\documentclass[preprint,aps,nofootinbib]{revtex4}
\usepackage{graphicx}
\usepackage{subfigure}
\usepackage{epsfig}
\usepackage{dcolumn}
\usepackage{bm}
\usepackage{ulem}
\usepackage{color}
\usepackage{multirow}
\usepackage{slashed}
\usepackage{hyperref}
\usepackage{amsthm}
\usepackage{amsmath}
\newcommand{\f}{\begin{equation}}
\newcommand{\ff}{\end{equation}}
\newcommand{\fa}{\begin{eqnarray}}
\newcommand{\ffa}{\end{eqnarray}}

\begin{document}
\title{Holographic superconductor induced by charge density waves}
\author{Yi Ling $^{1,2,4}$}
\email{lingy@ihep.ac.cn}
\author{Peng Liu $^{3}$}
\email{phylp@jnu.edu.cn}
\author{Meng-He Wu$^{1,2,4}$}
\email{mhwu@ihep.ac.cn}
 \affiliation{$^1$Institute of High
Energy Physics, Chinese Academy of Sciences, Beijing 100049,
China\\ $^2$ School of Physics, University of Chinese Academy of
Sciences, Beijing 100049, China\\ $^3$ Department of Physics and
Siyuan Laboratory, Jinan University, Guangzhou 510632, China \\
$^4$~ Shanghai Key Laboratory of High Temperature Superconductors,
Shanghai, 200444, China }

\begin{abstract}
Understanding the role of charge density wave (CDW) in
high-temperature superconductivity is a longstanding challenge in
condensed matter physics. We construct a holographic
superconductor model in which the $U(1)$ symmetry is spontaneously
broken only due to the presence of CDWs, rather than previously
known free charges with constant density. Below the critical
temperature of superconductivity, CDW phase and superconducting phase coexist, which is
also  justified by the numerical results of optical conductivity.
The competitive and cooperative relations between CDW phase and
superconducting phase are observed. This work supports
the opinion that the appearance of pseudo-gap in CDW phase
promotes the pre-pairing of electrons as well as holes such that
the formation of superconductivity benefits from the presence of
CDW.
\end{abstract}
\maketitle
\section{Introduction}
In the past decade, AdS/CMT duality
\cite{Hartnoll:2016apf,zaanen2015,ammon2015} has been becoming a
powerful tool for understanding the fundamental problems in
strongly coupled system since the seminal work of holographic superconductor
\cite{Gubser:2008px,Hartnoll:2008vx}. Considerable efforts have
been triggered to understand the mechanism of
high-temperature superconductivity from the holographic point of
view. The key insight in this approach is that the AdS
background may become unstable due to the presence of negative
mass terms such that the $U(1)$ gauge symmetry is spontaneously
broken below the critical temperature. A charged scalar
condensates and works as the order parameter to characterize the
superconducting phase in the boundary theory, following the
standard holographic dictionary.
In parallel, holographic models
of charge density waves (CDWs) have been constructed by
spontaneously breaking the translational symmetry in the bulk
\cite{Donos:2013gda} and its fundamental features as a
metal-insulator transition have been observed in
\cite{Ling:2014saa}. The strongly coupled nature of the above
systems has been signaled by the energy gap or pseudo-gap in the
optical conductivity which are much larger than that predicted by BCS theory \cite{Hartnoll:2008vx,Ling:2014saa}.

High temperature superconductor exhibits very abundant phase
structure, which involves the interplay between CDW phase and
superconducting phase. Disclosing their relation is crucial
for understanding the mechanism of high temperature
superconductivity. For cuprate oxides, the interplay between the
CDW phase, also known as the pseudo-gap phase, and the
superconducting phase has been under debate for decades (For
instance, see \cite{Vojta:2009} for review). The coexistence and the competition between the CDW phases and the superconducting phases are well-known \cite{Tranquada:1997,Hanaguri:2004,Calandra:2011,Denholme:2017,Gabovich:2002,Borisenko:2009,Eduardo H:2014}. Specifically, the competitive relation is signaled by the suppression of $T_c$ by the CDW phase \cite{Fujita:2002,Croft:2014}, and the competition of their order parameters \cite{neto:2013,chang:2012do}. Superconducting phase can grab charge carriers from CDW phases \cite{chang:2012do}. On the other hand, recent experiments reveal a novel cooperative relationship between them as well. This is signaled by the positive correlation between their critical temperatures \cite{Rahn:2012,comin:2014s,cho:2018cs}. The CDW assists the superconductivity through phonons which serve as a ``glue"  between  electrons to form a Cooper pair \cite{cho:2018cs}. Moreover, the CDW is argued to cause the superconductivity under certain conditions \cite{Castellani:1995,Tabis:2014}. Nonetheless, at present a complete theoretical interpretation on their relations has not been emerged in condensed matter physics.
One challenge comes from the strongly coupled nature of the high temperature superconductivity and the CDW phases (Recent experiments suggest that the mechanism of CDW phase in superconductors may involve strongly correlated physics \cite{flicker:2015nc,Chatterjee:2015nc,Ugeda:2016}.). This fact renders holographic duality theory an effective weapon for attacking them.

High temperature superconductivity is related to both free charge carriers and CDW.
To reveal the role of CDW solely in high temperature superconductivity, one good strategy is to separate it from free carriers in the compound. This is a hard task because typical high temperature superconductors are produced in the presence of both free charge carriers and CDW. In this paper, we construct a novel holographic superconductor model in
the presence of CDWs only. Previously, all the holographic
superconductor models are built in the presence of free charges
with constant density \cite{Arean:2013mta,Montull:2012fy},  including some models with CDW\cite{Kiritsis:2015hoa,Cremonini:2016rbd}\footnote{Even in the probe limit where the background is Schwarzschild-AdS black hole,
accompanying the condensation of scalar field, free charges with
constant density is perturbatively generated as well.}. For the
first time, our model clearly demonstrates that the
superconductivity may form from the pre-exiting CDW phase
 by a first order phase transition. This
may provide a direct evidence to the argument that the CDW state or the pseudo-gap state, is the precursor to the
superconducting states
\cite{Geshkenbein:1997,Kanigel:2008,Yang:2008,Kordyuk:2015,Seo:2019},
which has also been verified by the experiment on Nernst effect
\cite{Wang:2005}. Moreover, below the transition
temperature, our model is characterized by the coexistence of the
CDW phase and superconducting phase, which has been observed in
many superconducting materials as discovered in \cite{Denholme:2017,Hanaguri:2004,Calandra:2011,Tranquada:1997}. Furthermore, in our model the CDW phase can not
only compete with superconducting phase, but also cooperate with
superconducting phase under certain conditions, similar to the
phenomena observed in \cite{cho:2018cs}. Specifically,
the competitive relation is manifestly disclosed by the order parameters. Above the
critical temperature $T_c$ of superconductivity, the CDW order
parameter increases with the decrease of the temperature. However,
when the temperature drops below $T_c$, the order parameter of CDW
decreases while that of superconductivity increases, which exactly
coincides with the phenomena observed in
\cite{chang:2012do,neto:2013}. On the other hand, the positive correlation between their critical temperatures above a critical momentum mode suggests a cooperative relationship between them, which also coincides with experiment \cite{cho:2018cs}.
In addition, the system in our model is neutral in the sense that the average value of charge
density over a period vanishes, implying that the charge
excitations consist of both electron pairs and hole pairs.
The hole superconductivity as a promising candidate of high
temperature superconductivity was previously investigated in
\cite{Hirsch:2002tc,Hirsch:1986hole,Hirsch:2001prl}.

\section{The holographic setup}
In this section, we build the holographic model in the framework of
Einstein-Maxwell-Dilaton theory. The action reads as,
 \begin{equation}\label{eq:eps+1}
  \begin{aligned}
  S=&\frac{1}{2\kappa^2}\int d^4 x \sqrt{-g} \left[R-\frac{1}{2}\left ( \nabla \Phi  \right )^2-V\left ( \Phi  \right ) -\frac{1}{4}Z_{A}(\Phi )F^2  -\frac{1}{4}Z_{B}(\Phi )G^2\right. \\
 & \left.-\frac{1}{2}Z_{AB}(\Phi )FG - \left |\left ( \nabla  -ieB  \right )\Psi \right |^2-m_{v}^{2}\Psi \Psi^*\right],
  \end{aligned}
\end{equation}
where $F=dA$, $G=dB$, $Z_{A}(\Phi )=1-\frac{\beta }{2} L^2 \Phi
^2$, $Z_{B}(\Phi )=1$, $Z_{AB}(\Phi )=\frac{\gamma }{\sqrt{2}} L
\Phi $, $V ( \Phi )=-\frac{1}{L^{2}} + \frac{1}{2}m_{s}^{2}\Phi
^{2}$.  In this model, two $U(1)$ gauge fields are involved. Gauge
field $A$ is introduced to form an AdS-RN background with finite
Hawking temperature, while gauge field $B$ is treated as the
Maxwell field, whose dual is the electric current of the boundary system.
In addition, two scalar fields are
introduced. One is the real dilaton field $\Phi$ which is viewed as
the order parameter of translational symmetry breaking; while the
other complex field $\Psi$, charged under the gauge field
$B$, is viewed as the order parameter of $U(1)$ gauge symmetry
breaking. The term with parameter $\gamma$ is introduced to induce
the instability of homogeneous background so as to generate CDWs, while the term with parameter $\beta$ is to obtain a non-zero critical wavelength $k_c$ for CDW.

In the absence of the complex field $\Psi$, the holographic CDW as
well as its optical conductivity has previously been obtained within the above framework in \cite{Ling:2014saa}. After the spontaneous breaking of translational symmetry, the background with
CDW becomes spatially modulated. Without loss of generality, throughout this paper we set the AdS radius $l^2=6L^2=1/4$, the masses of the dilaton and the condensation $m_{s}^{2}=m_{v}^{2}=-2/l^2=-8$, the coupling constant $\beta=-130$ and $\gamma=16.6$,
then the maximal critical temperature for translational symmetry breaking is $T_{CDW}
\approx 0.071 \mu$ with the critical momentum mode $k_{c}\approx
0.335 \mu$, as illustrated in Fig.\ref{fig1}. In addition, the large $N$ limit of holographic duality requires that $\frac{l^2}{\kappa^2}\gg 1$.

The charge density can be read off from the expansion
$B_t =-\rho(x)z+O(z^2)$ with
\begin{eqnarray}\label{eq:eps+6}
 \rho(x)=\rho_0+\rho_1 \cos (k_c x)+...+\rho_n\cos (n k_c x).
 \end{eqnarray}
One remarkable feature of this two-gauge-field model is that all
the even-order coefficients of the charge density vanish, namely,
$\rho(x)=\rho_1 \cos (k_c x)+\rho_3\cos (3 k_c x)+\cdots$, which has
been justified by numerical analysis in \cite{Ling:2014saa}. It
means that the dual boundary theory contains only CDWs, without
free charges, namely, $\rho_0=0$. Thanks to this feature, the
Peierls phase transition as a typical metal-insulator transition
can be manifestly observed \cite{Ling:2014saa}.
This advantage persists even after the superconducting phase transition, as we will describe below.

\section{The instability and the critical temperature of superconductivity}
In this section, we examine the instability of striped black
brane with CDW and evaluate the critical temperature $T_{c}$ for the condensation of $\Psi$. At this stage we may set $\Psi$ as $\eta e^{i\theta
}$ and fix the gauge with $\theta=0$.

\begin{figure}[h]
  \includegraphics[width = 0.5\textwidth]{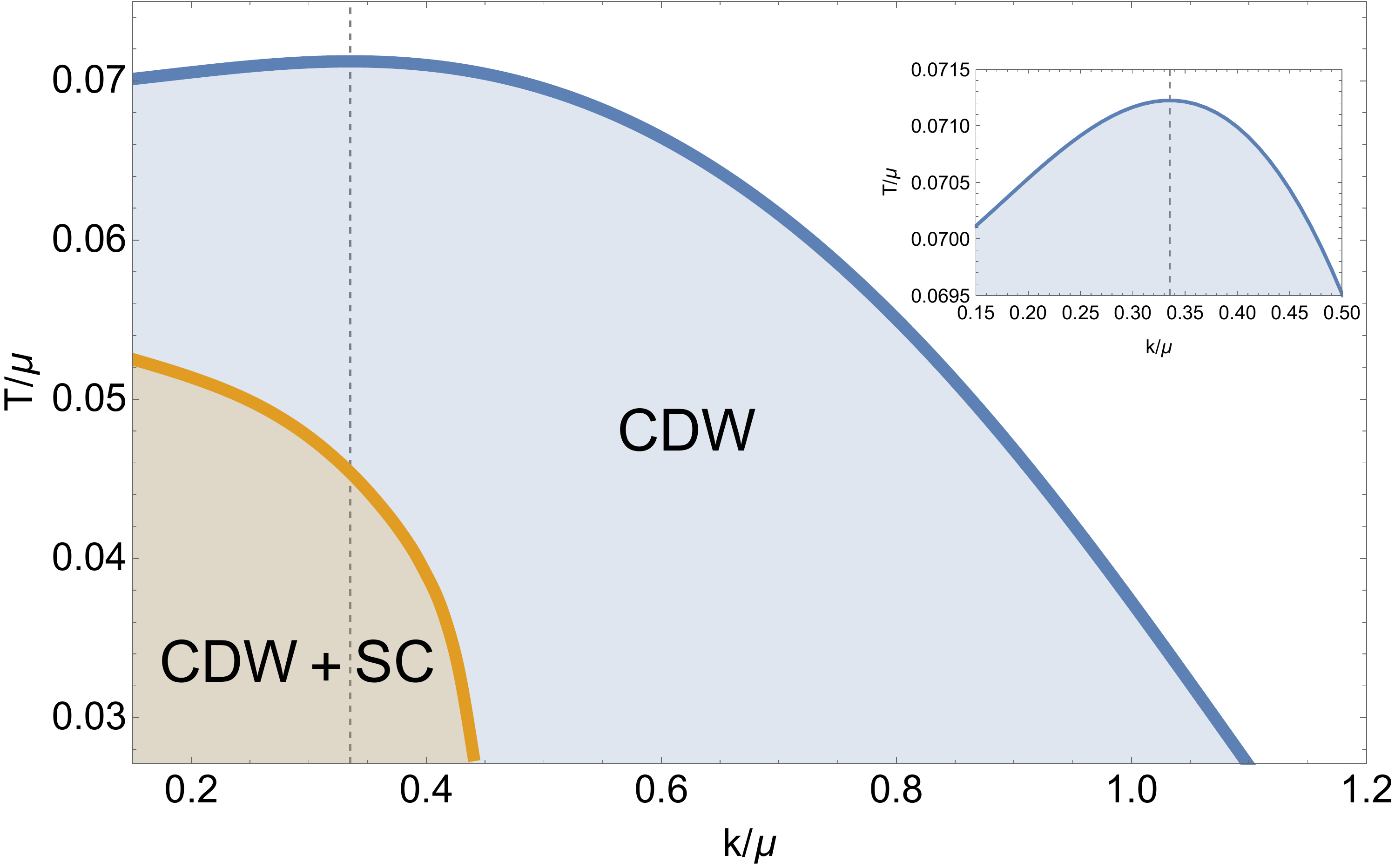}
  \caption{\label{fig1}
    The phase diagram in the plane of $(T/\mu, k/\mu)$. The blue curve denotes the critical temperature of CDW namely $T_{CDW}$ and the orange curve
    denotes the critical temperature of superconductivity, namely $T_{c}$. The inset zooms in the critical point of the CDW curve, and the dashed vertical line represents $k=k_c$.}
\end{figure}

As pointed out in \cite{Horowitz:2013jaa, Ling:2014laa}, whether
the background is stable or not can be treated as a positive
self-adjoint eigenvalue problem for $e^2$. Thus we rewrite the
equation of motion for $\eta$ as,
\begin{eqnarray}\label{eq:eps+7}
(\nabla  ^{2} - m_{v}^2) \eta = e^2 B^2\eta.
 \end{eqnarray}
The key condition for a non-zero $\eta$ solution is that
$B^2=g^{tt}B_tB_t \neq 0$. In all the previous holographic models,
this condition is guaranteed by the presence of non-zero charge
density $\rho_0$. Here, only CDW
presents and we wonder if it would induce the instability of the
background. The affirmative answer can be obtained by numerics. We search for non-zero solutions of $\eta$ by changing the value of charge $e$ (See Fig.\ref{fig2} in appendix \ref{appendix:b} ). Here we plot the phase diagram in Fig.\ref{fig1} with $e=4$ \footnote{For other
values of $e$, the phase diagram is qualitatively the same, except
that the coexisting region becomes larger or smaller, which can
also be justified in Fig.\ref{fig2} in appendix \ref{appendix:b}.}. First of all, the superconducting phase can only exist in the presence of CDW. This shows that CDW plays a crucial
role in the formation of superconducting phase. Another important phenomenon is the correlation between the critical temperature of
CDW phase and superconducting phase. For $k<k_c$, $T_{CDW}$  is
negatively correlated with $T_c$ (competitive relation); while for
$k>k_c$, $T_{CDW}$ is positively correlated with $T_c$
(cooperative relation). This novel relation is similar to
results of very recent experiments \cite{cho:2018cs}.

\section{Numerical solutions for background with superconductivity}
\begin{figure} [h]
  \center{
    \includegraphics[width = 0.5\textwidth]{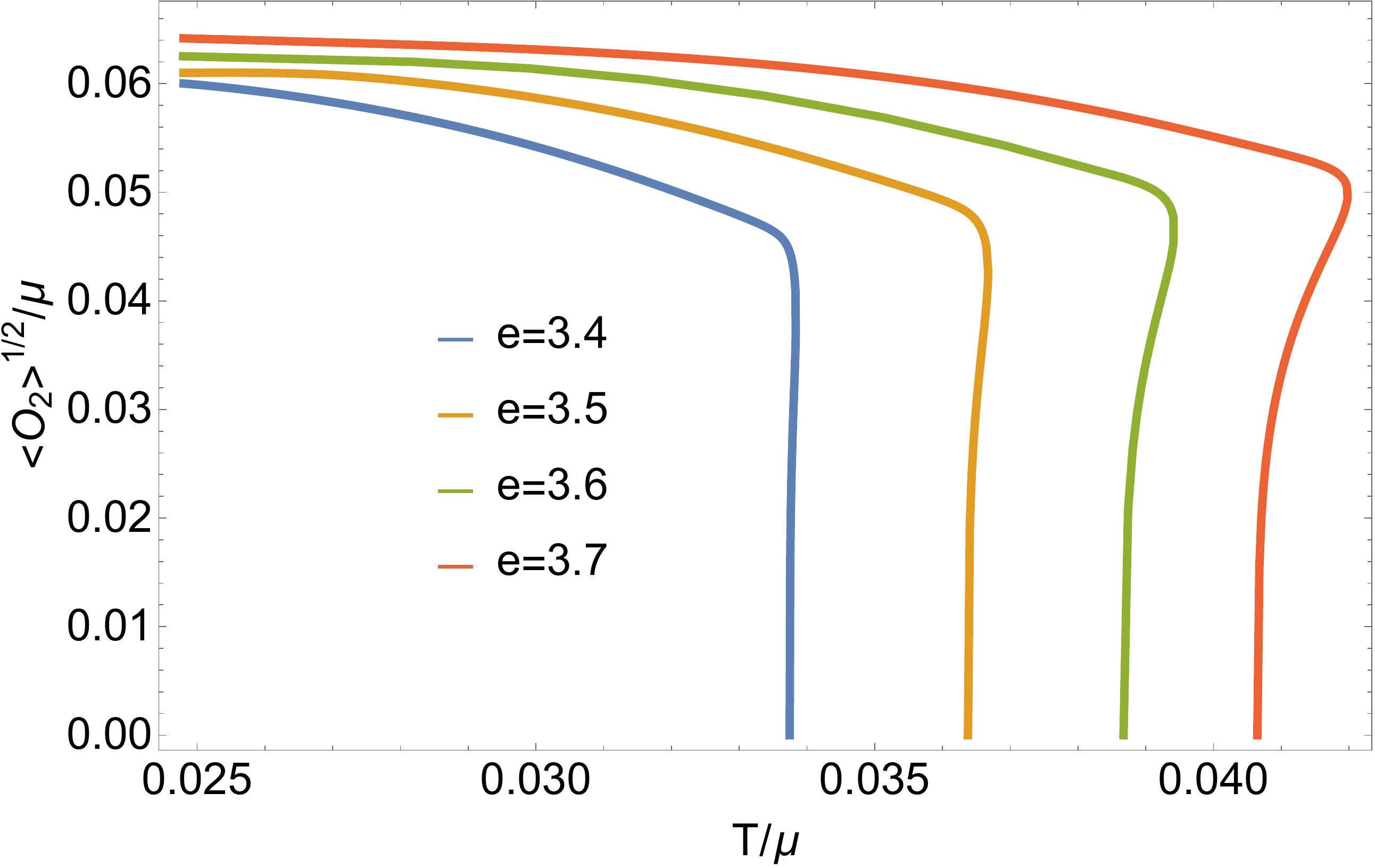}
    \caption{\label{fig3}
      The condensation of the charged scalar field for different values of $e$ at $k=k_c$.}
  }
\end{figure}

Next, we numerically solve the full holographic system.
Adopting the ansatz in appendix \ref{appendix:b}, the
resultant equations of motion consist of nine partial differential equations with respect to $x$ and $z$, which can be numerically solved with Einstein-DeTurck method
\cite{Headrick:2009pv,Horowitz:2012ky}. Especially, the near boundary expansion of $\eta$ is,
\begin{equation}\label{eq:eps+8}
  \begin{aligned}
    \eta=z \eta_1(x)+z^2 \eta_2(x) + \mathcal O(z^3).
  \end{aligned}
\end{equation}
We treat $\eta_1(x)$ and $\eta_2(x)$ as the source and expectation, respectively.
The $U(1)$ symmetry is expected to be broken spontaneously,
hence we set $\eta_1(x)=0$. Below $T_c$, a non-trivial
solution of the condensation term $\eta_2(x)$ can be expanded as,
\begin{equation}\label{eq:eps+9}
  \begin{aligned}
    \eta _{2}(x) = \eta _{2}^{(0)} +\eta _{2}^{(1)} \cos(k_{c}
    x)+\cdots+\eta _{2}^{(n)} \cos(nk_{c} x).
  \end{aligned}
\end{equation}
Numerically, we find that only even orders survive \footnote{Mathematically we
also find the other branch of solutions with odd orders only. But
these solutions are not physical, because $\eta_2$, as the
modulus of the condensation, must be positive-definite.}.
In Fig.\ref{fig3}, we plot the constant term $\eta^{(0)}_2$ of the condensation
vs $T$ at $k=k_{c}$. The condensation saturates in low temperature
region, while in the vicinity of critical temperature, the
condensation becomes a multi-value function of the temperature.
This feature clearly indicates a first-order phase
transition, which can also be justified with the analysis of the free energy.
 We found that, the branch of solution with higher condensation has lower free energy, that is favored by thermodynamics.

Next, we study the behavior of charge density and CDW with temperature after the superconducting phase transition. Surprisingly, we find that all the even orders of the charge density in Fourier expansion still vanish after condensation. Namely, it takes the form as $\rho(x)=\rho_1 \cos (k_c x)+\rho_3\cos (3 k_c x)+\cdots$. This means that the superconducting phase forms from CDWs, not from free charges.

\begin{figure} [h]
  \center{
    \includegraphics[width = 0.4\textwidth]{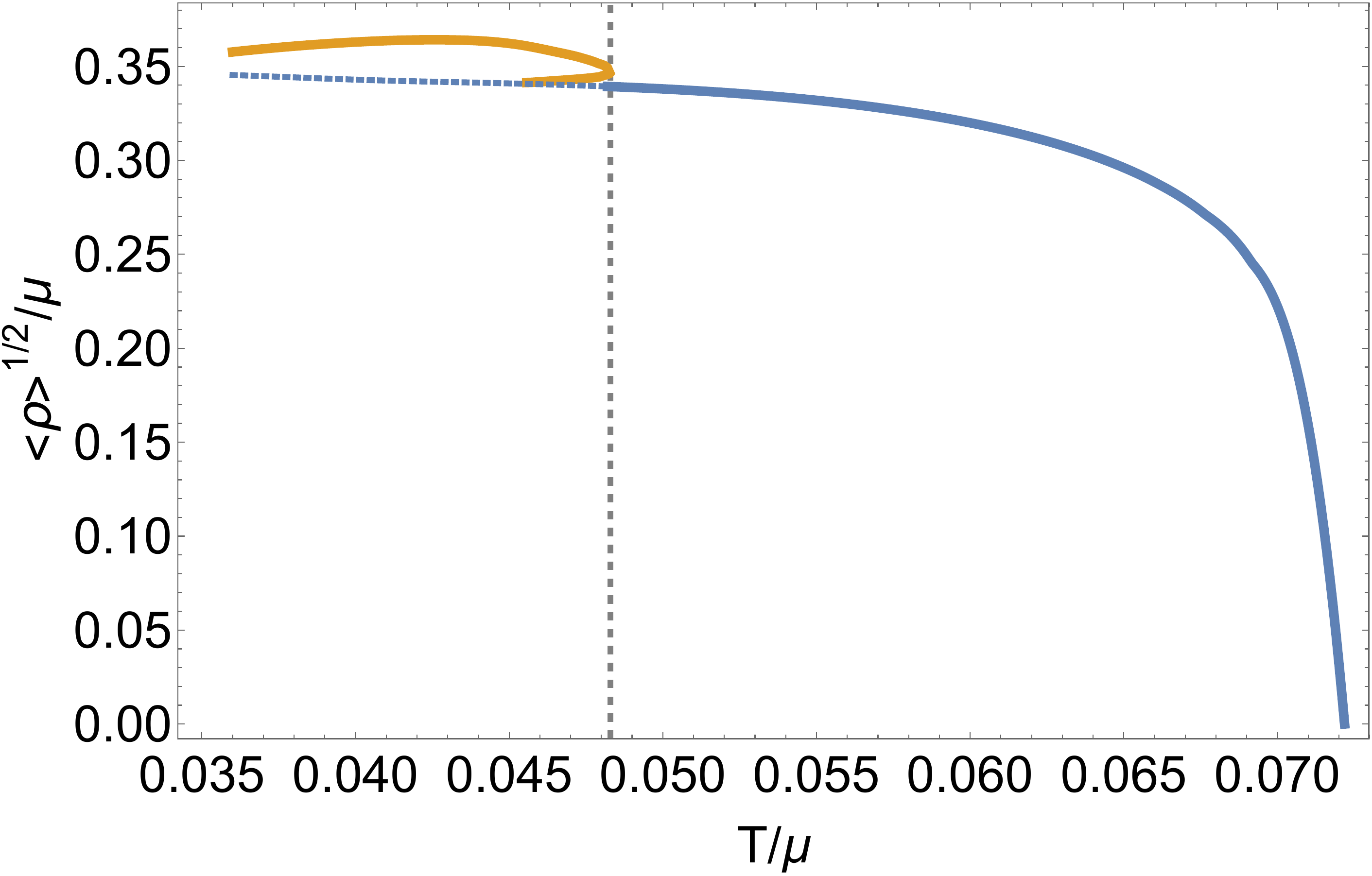} \quad \quad
    \includegraphics[width = 0.4\textwidth]{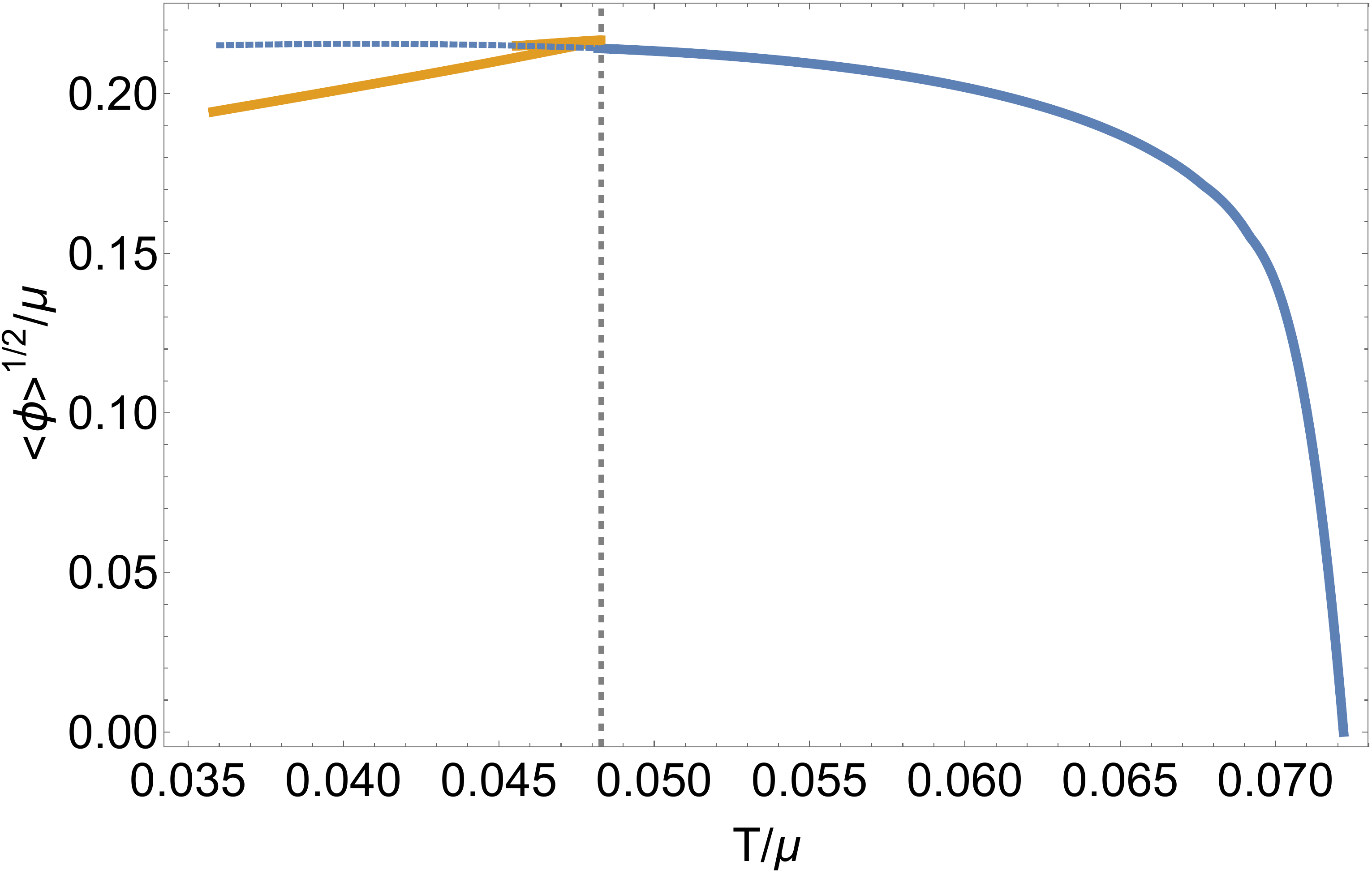}
    \caption{\label{fig4} The charge density $\rho_1$ and order parameter $\phi^{(1)}_{1}$ as functions of the temperature during the course of condensation.
      The vertical dotted line denotes the location of $T_c\approx 0.048 \mu$. Below $T_c$, the yellow curve is plotted with superconducting condensation, while the dotted curve is plotted without superconducting condensation.}}
\end{figure}

The near boundary expansion of the CDW order parameter is $\Phi=\phi_{1}(x)z^2 + \mathcal O(z^3)$, and we numerically find $\phi_{1}$ behaves as $\phi^{(1)}_{1} \cos(k_c x)+\phi^{(3)}_{1}\cos (3 k_c x)+\cdots$.
In Fig.\ref{fig4}, we plot $\rho_1$ and $\phi^{(1)}_{1}$ as
functions of the temperature with $e=4$. The left plot shows that after superconducting phase
transition, the charge density becomes larger in comparison with
that in the absence of superconducting phase transition. While
the right plot shows that the order parameter of
CDW is always smaller than that in the absence of superconducting phases.
This evidently indicates that the CDW is suppressed by the presence of the superconductivity.

The physics behind these novel phenomena can be understood as
follows. First of all, during all the process, no free charges
involve in but only CDWs present in the charge density. We
conclude that the superconductivity is induced by CDWs, rather than the free charge. This fact can also be justified by Eq.(\ref{eq:eps+7}).
Moreover, the increase of charge density after superconducting phase transition
indicates that the periodic charge density consists of both normal
carriers and superconducting carriers. Consequently, the carrier pairs
of superconducting phase must contain both charge pairs and hole
pairs, because the net charge is zero. This unconventional
mechanism of carrier pairing, such as hole superconductivity, is a
promising candidate for the mechanism of high temperature
superconductivity \cite{Hirsch:1986hole,Hirsch:2001prl,Hirsch:2002tc}. They beat the
traditional BCS theory in explaining the high frequency absorption
behavior and high transition temperature of superconductivity.

\section{The optical conductivity}
To justify that the above condensation due to $U(1)$ symmetry breaking gives rise to a superconducting state indeed, we study the optical conductivity by linear perturbations (see appendix \ref{appendix:c} for details). Here we only consider the linear response of $B$ field along $x$ direction, namely $\delta B_{x}$.
The optical conductivity can be read off from $\delta B_{x}=(1+j_{x}(x)z+...) e^{-i\omega t}$ as,
\begin{equation}\label{eq:accond}
  \sigma ( \omega ) =4 \frac{ j^{(0)}_{x}}{i \omega } ,
\end{equation}
where $j^{(0)}_{x}$ is the leading term of the Fourier expansion.

\begin{figure} [h]
  \center{
    \includegraphics[width = 0.4\textwidth]{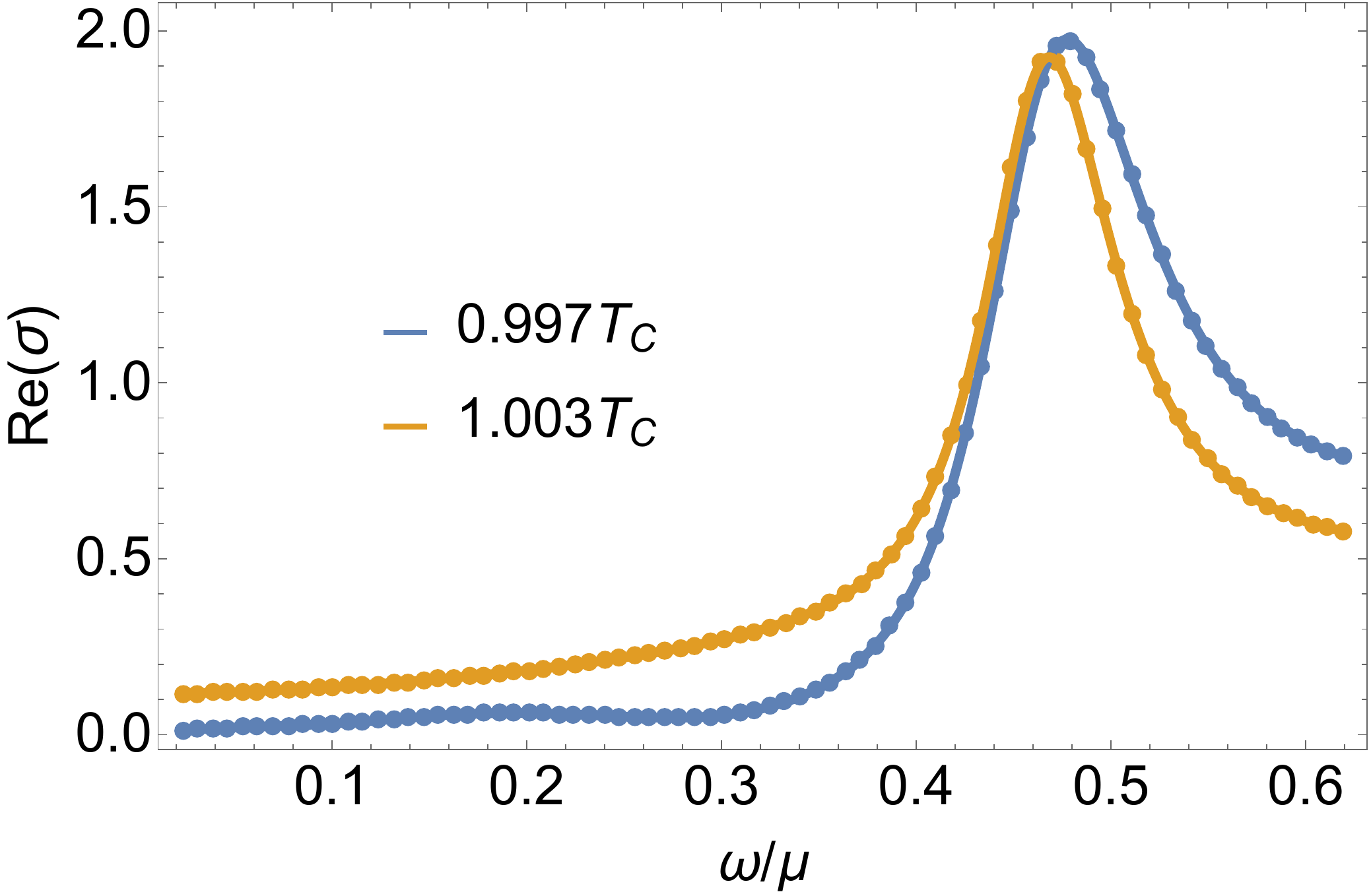} \quad \quad
    \includegraphics[width = 0.4\textwidth]{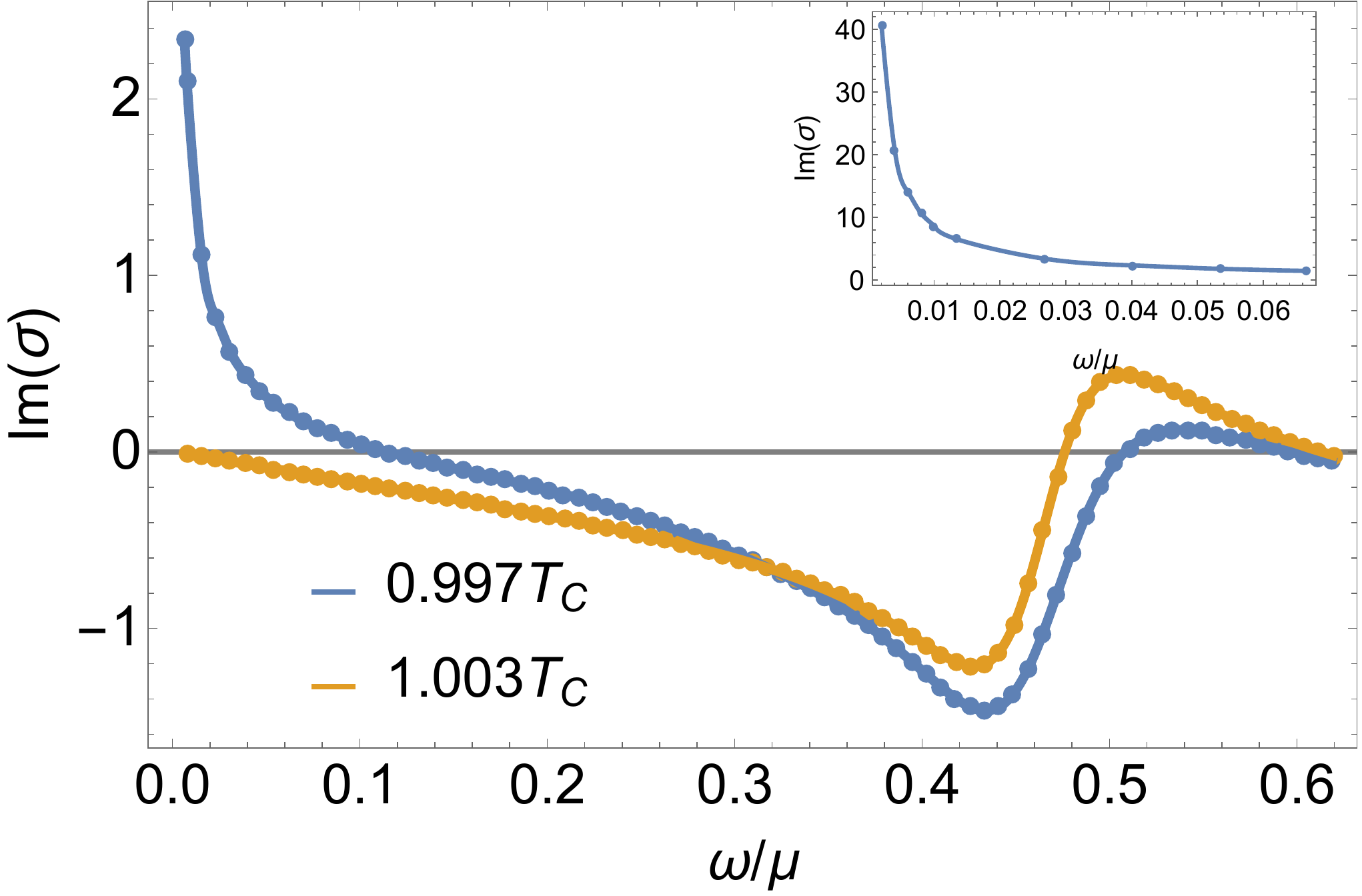}
    \caption{\label{fig5}
      The optical conductivity before and after the superconducting phase transition.
      The left plot is the real part of the optical conductivity and the right plot is the imaginary part.}
  }
\end{figure}

We plot the optical conductivity at temperatures close to $T_c$ in Fig.\ref{fig5}. Above $T_c$, the appearance of the pseudo-gap in the real part shows that the system is in the insulating phase, as revealed in
\cite{Ling:2014saa}, and the imaginary part of the conductivity converges to zero. We stress that the DC conductivity here is finite and approaches to zero as the temperature goes down from $T_{CDW}$,  which results from the vanishing average charge density. No charge is coupled with the goldstone modes in our model, hence no delta function in DC conductivity, which is in contrast to the cases in \cite{Amoretti:2017frz,Alberte:2017oqx,Amoretti:2017axe,Ammon:2019wci} where the goldstone modes carry non-zero charges. However, below $T_c$, the imaginary part exhibits a power law $\text{Im}\,\sigma \sim 1/\omega$ in low frequency region, implying that a delta function arises at $\omega=0$ in the real part of the conductivity by Kramers-Kronig relations. It indicates that the
system undergoes a phase transition from insulating phase to superconducting phase indeed.

In low frequency region, we fit the optical conductivity with the following formula
\begin{equation}\label{eq:eps+12}
  \sigma (\omega )  =i \frac{K_{s}}{\omega } + \frac{K_{1} \tau_{1} }{1-i\omega \tau (1-\omega^{2} _{01}/\omega^{2})} + \frac{K_{2} \tau_{2} }{1-i\omega \tau (1-\omega^{2} _{02}/\omega^{2})},
\end{equation}
where $K_{s}$ is proportional to the number of superfluid density, and $K_{i}$ is proportional to the number density of CDW, while $\tau_{i}$ is
the relaxation time, and $\omega _{0i}$ is the average
resonance frequency ($i=1,2$). The fitted parameters are
listed in Table. \ref{tab1}. From this table, one can see that
the superfluid density becomes non-zero below the critical temperature $T_c$.

\begin{table}[ht]
  \begin{tabular}{| p{2cm}<{\centering}| p{2cm}<{\centering}| p{2cm}<{\centering}| p{2cm}<{\centering}| p{2cm}<{\centering}| p{2cm}<{\centering}| p{2cm}<{\centering}| p{2cm}<{\centering}|}
    \hline
    $T/T_c$ & $K_{s}/\mu $ & $K_{1}/\mu $ & $\tau _{1}\mu$ & $\omega _{01}/\mu$ & $K_{2}/\mu $ & $\tau _{1}\mu$ & $\omega _{02}/\mu$ \\ \hline
    $1.007$ & $0$          & $0.136$      & $12.240$       & $0.468$            & $3.279$      & $0.302$        & $1.794$            \\ \hline
    $0.997$ & $0.020$      & $0.159$      & $9.947$        & $0.483$            & $2.893$      & $0.431$        & $1.457$            \\ \hline
  \end{tabular}
  \caption{\label{tab1}
    The fitted parameters for optical conductivity.
  }
\end{table}
In the end, we remark that the energy gap of the superconductor is almost the same as the pseudo-gap of CDW, $\Delta_{CDW}\approx\Delta_c\approx 9.041 T_c$, both of which are much larger than the BCS value. The above phase transition implies that the appearance of pseudo-gap can promote the pre-pairing of carriers and support the cooperative relation between the pseudo-gap phase and the superconducting phase. Similar phenomenon was previously observed in a holographic model with novel insulator as well \cite{Ling:2017naw}.

\section{Discussion}
We have constructed a novel holographic model in which the role of CDW during the phase transition of superconductivity has been clearly disclosed. First of all, the $U(1)$ gauge symmetry in the bulk is spontaneously broken due to the presence of CDW only, indicating that the superconductivity can form from the pre-existing CDW phase by the first-order phase transition.  Secondly, below the critical temperature $T_c$, the system is characterized by the coexistence of the CDW phase and the superconducting phase. It is found  that CDW phase and superconducting phase can both cooperate and compete with each other. Furthermore, the system remains neutral during the phase transition, implying that the charge excitations are composed of both electron pairs and hole pairs, beyond the traditional Cooper pairs. It also supports the opinion that  the pseudo-gap in CDW phase promotes the pre-formed pairs of carriers such that the formation of superconductivity benefits from the presence of CDW. Our work has  provided a novel
understanding on the complicated relationship between CDW and superconducting phases from the holographic  point of view, and may shed light on the mechanism of high temperature superconductivity.

The most desirable work next is to further investigate the features of superconductivity induced by CDW, and compare them with those induced by free charges, in particular, their responses to external magnetic field or disorder effects. Moreover, it is intriguing to investigate the relation between CDW and superconductivity in more complicated environment, since the practical high temperature superconductors are doped compounds with free carriers. To this end, one may turn on the chemical potential and the free charges with constant density such that the black hole background is charged under the gauge field $B$ as well. In this situation, the superconductivity may be induced by both CDW and free charges, analogous to the materials in experiments. Furthermore, the order parameters of CDW and superconductor are not directly coupled in current model. We definitely can introduce the coupling of these order parameters to explore the complicated relations between CDW and superconductors. Last but not the least, one may treat gauge field $A$ as the doping and consider its impacts on the relations between CDW and superconductors.

\section*{Acknowledgments}

We are very grateful to Matteo Baggioli, Yuxuan Liu, Chao Niu, Zhuoyu Xian and Yikang Xiao for helpful
discussions and suggestions. Y.L. would also like to thank Weijia Li, Hong Liu, Yu Tian, Jianpin Wu, Junbao Wu, Hongbao Zhang for previous collaboration or discussion on relevant topics. This work is supported by the Natural
Science Foundation of China under Grant No. 11575195, 11875053, 11905083 and 11847055.

\appendix

\section{The equations of motion} \label{appendix:c}

In this appendix, we derive the equations of motion for our model. We rewrite the $U(1)$ charged complex scalar field $\Psi$ as $\eta e^{i\theta }$, where $\eta>0$ is a real scalar field and $\theta$ is a St\"uckelberg field, then the action can be cast as,
\begin{equation}\label{eq:eps+2}
  \begin{aligned}
    S= & \frac{1}{2\kappa^2}\int d^4 x \sqrt{-g} \left[R-\frac{1}{2}\left ( \nabla \Phi  \right )^2-V\left ( \Phi  \right )-\frac{1}{4}Z_{A}(\Phi )F^2 -\frac{1}{4}Z_{B}(\Phi )G^2  \right. \\
       & \left. -\frac{1}{2}Z_{AB}(\Phi )FG -\left ( \nabla \eta   \right )^2-m_{v}^{2}\eta  ^{2} - \eta ^{2} (\nabla \theta - e B)^{2}\right].
  \end{aligned}
\end{equation}
The equations of motion for all fields can be written as,
\begin{equation}\label{eq:eps+3}
  \begin{aligned}
     & R_{\mu \nu }-T_{\mu \nu }^{\Phi }-T_{\mu \nu }^{A }-T_{\mu \nu }^{B }-T_{\mu \nu }^{AB }-T_{\mu \nu }^{\eta  } -T_{\mu \nu }^{\theta } =0, \\
     & \nabla  ^{2} \Phi -\frac{1}{4}Z_{A}' F^2 -\frac{1}{4}Z_{B}' G^2 -\frac{1}{2}Z_{AB}' FG -V'=0,                                              \\
     & \nabla  ^{2} \eta  - m_{v}^2 \eta  - (\nabla \theta -e B)^2\eta  =0,                                                                       \\
     & \nabla_{\mu }(Z_A F^{\mu \nu }+ Z_{AB } G^{\mu \nu })=0,                                                                                   \\
     & \nabla_{\mu }(Z_B G^{\mu \nu }+ Z_{AB } F^{\mu \nu })+2 e \eta ^2( \nabla^{ \nu}\theta  -  e B^{\nu } )=0,                                 \\
     & \nabla_{\mu } (\eta ^2 (\nabla^{\mu } \theta -e B^{\mu} ) )=0,
  \end{aligned}
\end{equation}
where
\begin{equation}\label{eq:eps+4}
  \begin{aligned}
    T_{\mu \nu }^{\Phi }   & =\frac{1}{2} \nabla_{\mu }\Phi \nabla_{\nu  }\Phi +\frac{1}{2}V g_{\mu \nu },                                       \\
    T_{\mu \nu }^{A  }     & =\frac{Z_{A}}{2} \left(F_{\mu \rho  }F^{\rho }_{\nu }- \frac{1}{4} g_{\mu \nu }F^2\right),                          \\
    T_{\mu \nu }^{B  }     & =\frac{Z_{B}}{2} \left(G_{\mu \rho  }G^{\rho }_{\nu }- \frac{1}{4} g_{\mu \nu }G^2\right),                          \\
    T_{\mu \nu }^{AB  }    & =Z_{AB}\left(F_{(\mu |\rho|  }G^{\rho }_{\nu ) }- \frac{1}{4}  g_{\mu \nu }FG\right),                               \\
    T_{\mu \nu }^{\eta  }  & = \nabla_{\mu }\eta \nabla_{\nu  }\eta  +e^{2}\eta ^{2}B_{\mu }B_{\nu } +\frac{1}{2}m_{v}^{2}\eta ^{2}g_{\mu \nu }, \\
    T_{\mu \nu }^{\theta } & = \eta ^{2} (  \nabla_{\mu  }\theta   \nabla_{\nu   }\theta  -2 e B_{ (\mu } \nabla_{\nu )  } \theta   ),
  \end{aligned}
\end{equation}
and the prime denotes the derivative with respect to ${\Phi }$.

\section{The numerical analysis of the background } \label{appendix:b}
We adopt the following ansatz for the background with CDW,
\begin{equation}\label{eq:eps+5}
  \begin{aligned}
     & ds^{2}=\frac{1}{z^2}\left[-(1-z)p(z)Qdt^{2}+\frac{Sdz^{2}}{(1-z)p(z)}+Vdy^{2} +T(dx + z^2 Udz)^2\right], \\
     & A=\mu(1-z)\psi dt, \ \  B=(1-z)\chi dt, \ \ \Phi =z\phi,  \ \ \eta =z \zeta   , \ \ \theta = 0,
  \end{aligned}
\end{equation}
where $p(z)=4 \left(1+z+z^2-\frac{\mu ^2 z^3}{16} \right)$ and
$\mu$ is the chemical potential of $A$ that is taken as the scaling unit
of the system. $Q$, $S$, $V$, $T$, $U$, $\psi$, $\chi$, $\phi$,
$\zeta  $ are functions of $x$ and $z$. Thanks to the
Einstein-DeTurck method, the Hawking temperature of the black
hole with stripes is simply given by $T/ \mu=(48-\mu^2)/(16 \pi \mu)$. Notice that if we set $Q=S=V=T=\psi=1$ and
$U=\chi=\phi=\zeta=0$, then the background goes back to the AdS-RN black hole.
\begin{figure} [h]
  \center{
    \includegraphics[width = 0.5\textwidth]{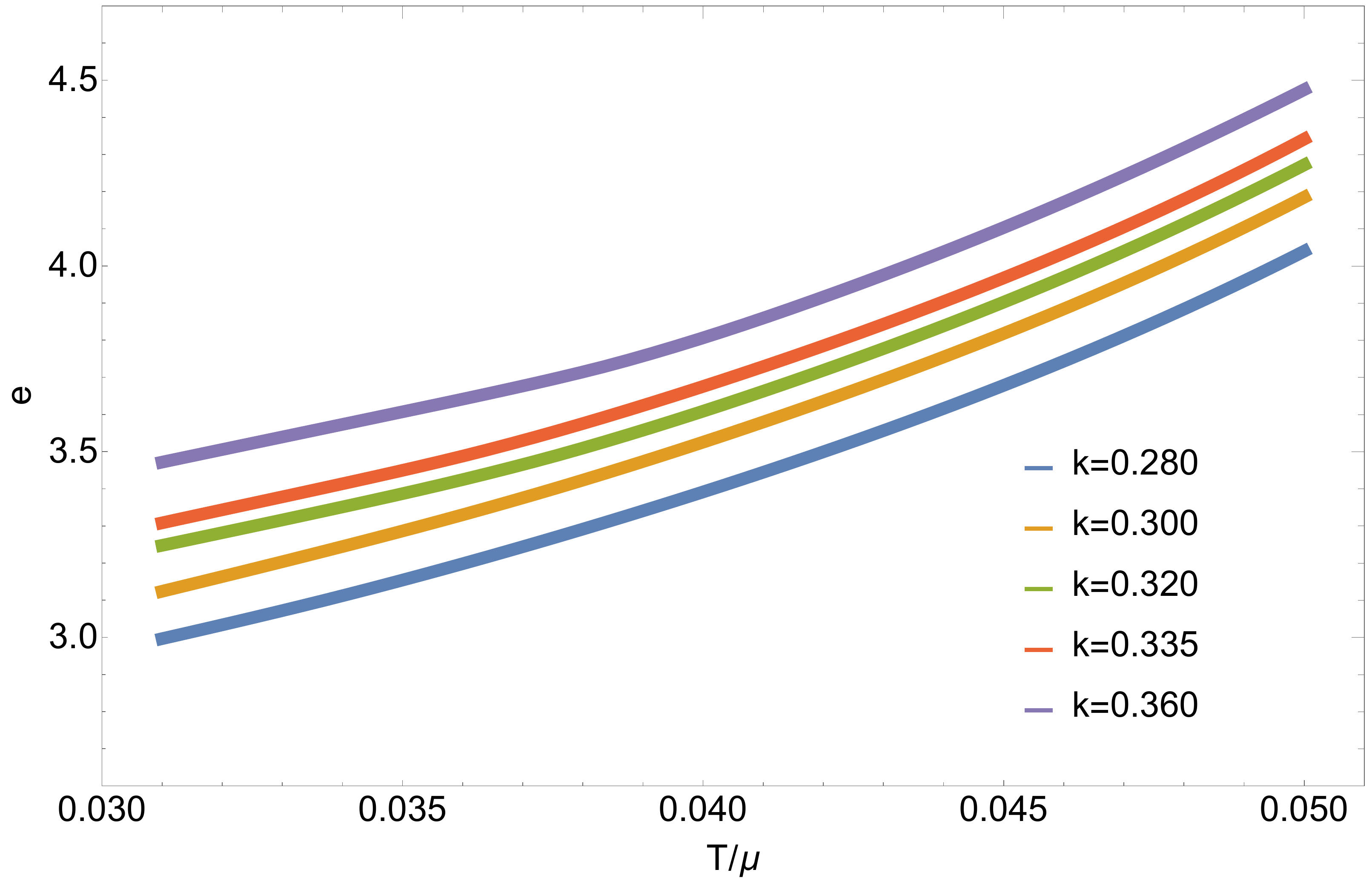}
    \caption{\label{fig2}
      The relation between the charge $e$ and the critical temperature $T_{c}$ for the condensate of the charged scalar field, where for each curve the momentum mode $k$ of CDW is fixed.}
  }
\end{figure}

First, we compute the critical temperature $T_c$ by turning on $\eta$ in probe limit and solving (\ref{eq:eps+8}). We plot the relation between the charge $e$ and $T_{c}$ in Fig.\ref{fig2}.
It indicates that below the critical temperature $T_{c}$,
the background with CDW becomes unstable and the complex scalar
hair starts to condensate. Moreover, the condensation become easier for larger charge, and for smaller momentum modes.

\begin{figure} [h]
  \center{
    \includegraphics[width = 0.4\textwidth]{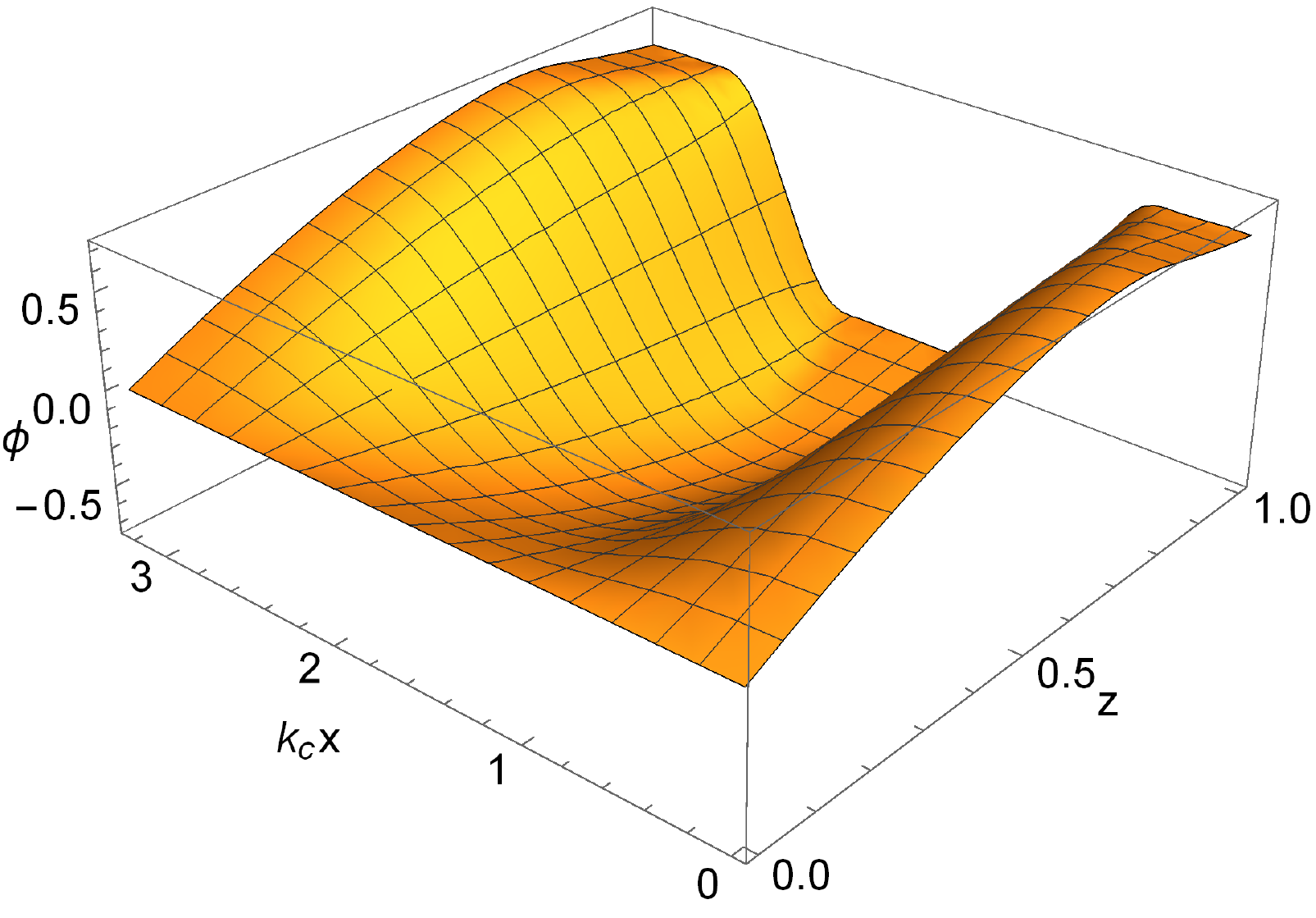} \quad \quad
    \includegraphics[width = 0.4\textwidth]{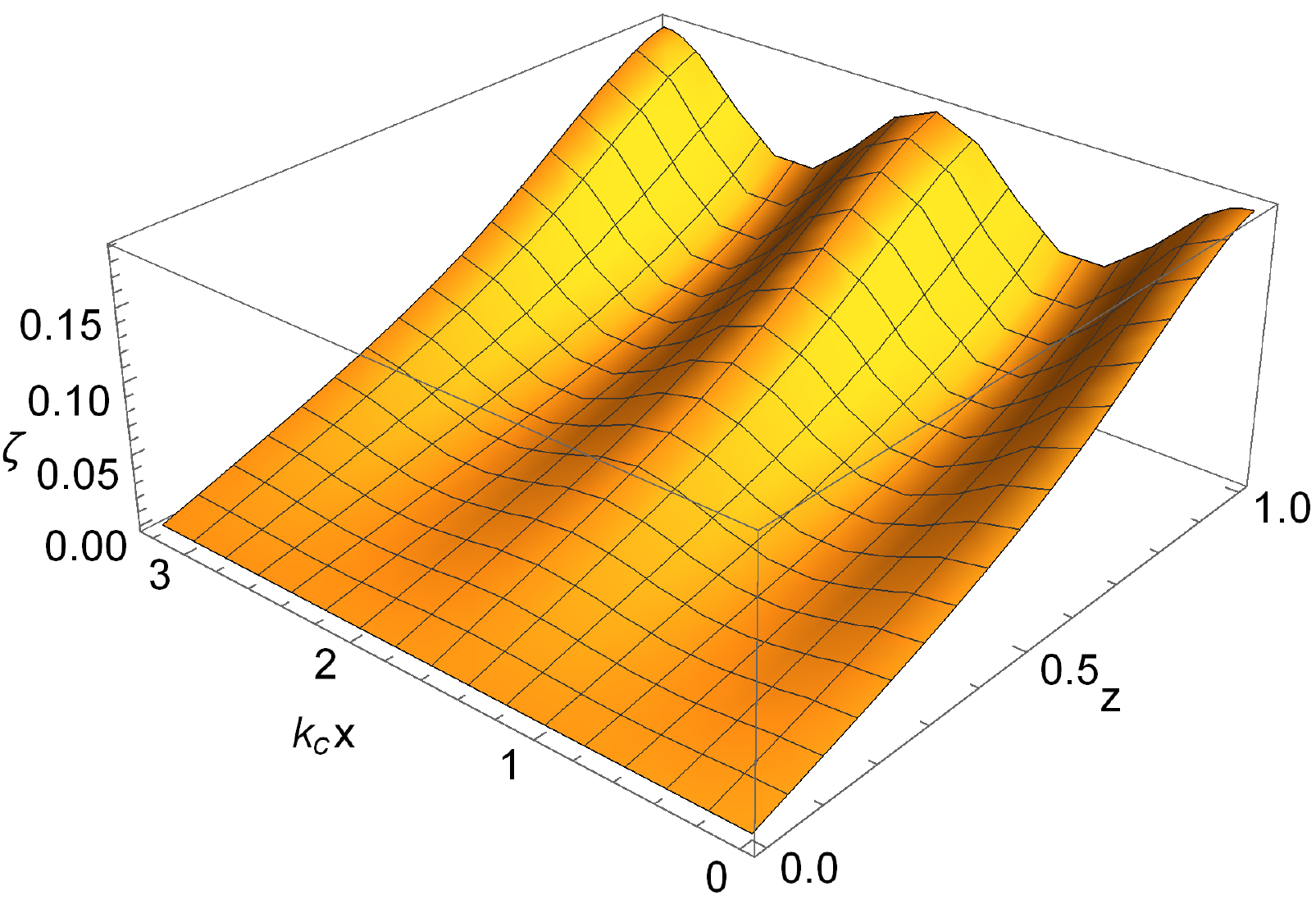}
    \caption{\label{fig6}
      The solution of the scalar $\phi$ and the charged scalar field $\zeta $ for $T=0.988T_c$, $e=4$ and $k=k_c$.
    }
  }
\end{figure}

Next, we numerically solve the equations of motion (\ref{eq:eps+3}) with the ansatz (\ref{eq:eps+5}). We apply Einstein-DeTurck
method to fix the coordinates with appropriate boundary conditions \cite{Headrick:2009pv}. In addition, we impose the regular boundary condition on the horizon, and require an asymptotic AdS$_4$ on the boundary.
With the equations of motion and boundary conditions, we obtain the numerical solutions by
the pseudo-spectral method and Newton-Raphson iteration
method \cite{Horowitz:2012ky,Horowitz:2013jaa,Ling:2014saa}.
As an example, Fig.\ref{fig6} shows the solutions for scalar field $\phi$ and the charged scalar field $\zeta$ at the temperature $T=0.988T_c$ with $e=4$ and $k=k_c$.

\section{The numerical analysis of the linear perturbations } \label{appendix:c}

Consider the following linear perturbation,
\begin{equation}\label{eq:eps+14}
  \begin{aligned}
    g_{\mu \nu } & =\bar{g}_{\mu \nu }+  \delta g_{\mu \nu }, & \ A_{\mu } & = \bar{A}_{\mu }+ \delta A_{\mu }, & \ B_{\mu } & = \bar{B}_{\mu }+ \delta B_{\mu }, \\
    \Phi         & =\bar{\Phi }+ \delta \Phi ,                & \zeta      & =\bar{\zeta  }+ \delta \zeta ,     & \theta     & =\bar{\theta }+ \delta \theta.
  \end{aligned}
\end{equation}
We mark the unperturbated quantities with an overbar in the above formula. Each perturbation oscillates with time as $e^{-i\omega t}$. In order to extract the optical conductivity along $x$ direction, we only turn on $B_x$ and keep $B_y = 0$.
Note also that we must turn on the St\"uckelberg field $\theta$ for self-consistency.
To fix the solutions, we adopt the Donder gauge condition and Lorentz gauge condition,
\begin{equation}\label{eq:eps+13}
  \bar{\nabla}^{\mu } \hat{h}_{\mu \nu }=0, \ \ \bar{\nabla}^{\mu } \delta A_{\mu }=0, \ \  \bar{\nabla}^{\mu } \delta B_{\mu }=0,
\end{equation}
where $\hat{h}_{\mu \nu }= \delta g_{\mu \nu }- \delta g
\bar{g}_{\mu \nu }/2$. As a result,
we obtain 21 linear equations of motion for the perturbations. We impose the ingoing boundary conditions on the horizon. Then the whole perturbation system can be solved by pseudo-spectral method.

\end{document}